\documentstyle[12pt]{article}
\newcommand{\argmin}{\mathop{\rm argmin}\limits}

\newcommand{\const}{\mathop{\rm const}\limits}

\newcommand{\grad}{\mathop{\rm grad}\limits}

\begin{document}
\begin{center}

{\bf OPTIMAL ADAPTIVE MULTIDIMENSIONAL-TIME SIGNAL ENERGY } \\

\vspace{2mm}

 {\bf ESTIMATION ON THE BACKGROUND NOISE.}\\

\vspace{2mm}

                Eugene Ostrovsky \\
{\it  Department of Mathematic, Soniclynx  company, \\
78042, Rosh Ha'ain, Hamelecha  street, 22, Israel; } \\
     e-mail: eugeny@soniclynx.net \\

\vspace{2mm}

                Eugene Rogover \\
{\it Department of Mathematics and Statistics, Bar \ - \ Ilan University,
59200, Ramat Gan, Israel.}\\
e \ - \ mail: \ rogovee@gmail.com \\

\vspace{2mm}
               Leonid Sirota \\
Department of  Mathematic, Bar \ - \ Ilan University,  Ramat Gan,
52900,   Ramat Gan, Israel, \\
 e-mail: sirota@zahav.net.il \\

\vspace{3mm}
                    {\sc Abstract.}\\

 \end{center}

 \vspace{3mm}

 We construct an adaptive asymptotically  optimal in the classical  norm of the space
 $ L(2, \Omega) $ of square integrable random variables the {\it Energy estimation }
of a signal (function) observed in some points (plan of experiment) on the background noise.\par

 \vspace{3mm}

 \par

\vspace{3mm}

 {\bf 1. Statement of problem. Notations.} Let  $ V(n),  n = 16,17, \ldots  $ be a sequence of a vector \ - \ valued  (in general case)
sets  (plans of experiences) in the cube $ [-\pi,\pi]^d, \ d = 1, 2,3, \ldots: $

$$
V(n) = \{ x_i = \vec{x}_i = \vec{x}_i(n), \}, \
\vec{x}_i \in [-\pi, \pi]^d.
$$

 At the points $ \vec{x}_i $ we observe the unknown signal (process, field)
$ f = f(x), \ x \in [-\pi,\pi]^d $ on the background noise:
$$
y(i) = f(\vec{x}_i) + \sigma \ \xi_i, \eqno(1)
$$
where the sequence noise $ \{ \xi_i \}, \ $ is the sequence of errors of measurements,
is the sequence of independent (or weakly dependent) centered: $ {\bf E } \xi_i = 0 $ normalized: $ {\bf Var }(\xi_i) = 1 $ random variables, $ \sigma =
\const > 0 $ is a standard deviation of errors. \par

 Let us denote by the $ E_m $ the so \ - \ called {\it Energy} operator, i.e. such that
 the Energy of the signal $ f $ may be defined as follows:
 (at last in the case $ m = 0,1,2 $  )

$$
W_m(f) \stackrel{def}{=} ||E_m \ f||^2L_2(-\pi,\pi)^d, \eqno(2)
$$
where we define for an arbitrary $ M = 1,2,\ldots \ - \ $ dimensional vector
$ z = (z_1, z_2, \ldots,z_M) \ |z| = \sqrt{\sum_{l=1}^M z_l^2 } $ and for the vector function $ g = g(x) = (g_1, g_2, \ldots, g_M) $
$$
||g||^2 L_2[-\pi, \pi]^d = || g ||^2 \stackrel{def}{=} \int_{(-\pi, \pi)^d } |g(x)|^2 \ dx, \ dx = dx_1 dx_2 \ldots \ dx_d. \eqno(3)
$$

Note that the functional

$$
||f||S(m) = \left[ \sum_{l=0}^m W_l(f) \right]^{1/2}
$$
is the classical Hilbert's - Sobolev's norm. \par

{\it We assume that the operator $ E_m = E_m(f) $ is the spherical invariant homogeneous differential operator of order } $ m, m = 0,1,2, \ldots. $ For example:

$$
E_0(f) = f; \ E_1(f) = \grad(f), \ E_2(f) = \Delta(f) = \sum_{l=1}^M
\frac{\partial^2 f }{\partial x_l^2}. \eqno(4)
$$

 Note that if $ m = 2 r $ is an even number, then $ E_m = \Delta^r.$ \par
 Note that our assumption (4) is true when the environment is homogeneous and isotropic.\par

\vspace{2mm}

{\bf Our goal is to offer and investigate an adaptive asymptotically
as $ n \to \infty $ optimal in the probabilistic $ L_2(\Omega) $ sense estimation (measurement) $ W_m(f,n) = W_m(n) $ of a energy functional} $ W_m(f) = W_m, $
{\bf based on the observations } $ \{ y(i), \ i = 1,2,\ldots,n \}. $ \par

\vspace{2mm}

 The probabilistic $ L_2(\Omega) $ sense denotes that we consider the following loss function:

 $$
 Z(W_m(f), W_m(f,n)) = {\bf E} \left[ W_m(f,n) \ - \ W_m(f) \right]^2. \eqno(5)
 $$
Here $ W_m(f,n) = W_m(n) $ denotes the estimation of the functional $ W_m = W_m(f), $
adaptive or not.\par

 The considered
 problem appears in the financial mathematic [4], technical diagnosis and geophysics [3], image processing [5] etc. \par

The one \ - \ dimensional case is consider in [1]\ - \  [3] etc.  We notice that there are some essential differences between the one \ - \ dimensional and multidimensional cases; we will show, for example, that in the multidimensional case we need to use only the  {\it optimal experience design.}  \par

  On the other words, this problem is called "filtration of a signal on the
background phone", "adaptive noise canceler" or "regression problem".\par
 In the one \ - \ dimensional case $ d = 1 $ this problem was considered in many publications ([1] \ - \ [5] etc). The case $ d = 2 \ $ is known as "picture processing" or equally "image processing".  \par

\vspace{2mm}

{\bf 2. Denotations. Assumptions. Construction of our estimations.} Let
$ \vec{z} = z = \{z_j \}, j = 1,2,\ldots, d, \ z_j \in [-\pi,\pi]  $  be a $ d \ - $ dimensional vector,
$$
F(\vec{z}) = (2\pi)^{-d} \prod_{j=1}^d (\pi + z_j),  \ \delta(n) =
\delta(n,V(n)) =
$$
$$
\sup_{z \in [-\pi, \pi^d] } |G_n(z) \ - \ F(z)|, \ G_n(z) = n^{-1} \sum_{i=1}^n I(x_i
< z),
$$
where
$$
I( \vec{x} < \vec{z}) = 1 \Leftrightarrow \forall j = 1,2, \ldots,d \
\Rightarrow \ x_j < z_j,
$$
and $ I(\vec{x} < \vec{z} ) = 0 $ in other case. \par
 The value, or more exactly, the function $ \delta = \delta(n) = \delta(n,V(n)) $  is called {\it discrepancy} of a sequence of the plans $  V(n). $ \par
 We suppose that

$$
\delta(n) = \delta(n, V(n)) \le  C(1,d) [\log( n)]^d /n, \eqno(6)
$$

 Note that in the one \ - \ dimensional case the condition (6) is
satisfied even without the member $ \log^d(n) $
if $ x_i = -\pi + 2 \pi i/n $ (the uniform plan); but in general case $ d \ge 2 $ we need to use, e.g., the classical Van der Corput or the
 Niederreiter's sequences  {\it (experience design)} \ (see [8], p. 183 \  -  \ 202), for which the condition (6) is satisfied.\par
 Note in addition that the Niederreiter's sequences allow us to elaborate  the
convenient for application {\it sequential } estimation of the energy of signal
 $ W_m(f). $ \par
 Recall that for arbitrary sequences of plans $ \{ V(n) \} $

$$
\delta(n) = \delta(n, V(n)) \ge  C(2,d) [\log( n)]^{d \ - \ 1} /n, n \ge 3.
$$



 We define  also $ \nu = 2^{1/d}, \  N_{d,m}(n) = 0.25 [n^{2/(4m + d )}] $
 and for $ N \in \ (1,  \ N_{d,m}(n)) $ the "rectangles"
$$
R(N) = \{ \vec{k}: \max_j k(j) \le N \},
 \ R_2(N) = R( [\nu \ N]) \setminus R(N),
$$

$$
R_3(N) = \{ \vec{k}: \min_j k(j) \ge N+1 \} = Z_+^d \setminus R(N).
$$
 Here  $ [z] $ denotes the integer part of a (positive) variable $ \ z.$\par

 Let us use the classical orthonormal trigonometric system of a functions defined on
 the circle $ [ -\pi, \pi]: $

$$
\psi_1(x) = (2\pi)^{-1/2}, \ \psi_2(x) = \pi^{-1/2} \sin x, \psi_3(x) = \pi^{-1/2}
\cos x, \psi_4(x) = \pi^{-1/2} \sin (2x),
$$

$$
\psi_5(x) = \pi^{-1/2} \cos (2x),\psi_6(x) = \pi^{-1/2} \sin(3 x), \psi_7(x) =
\cos(3x)
$$
etc. We define for the multivariate {\it integer} index
$ k = \vec{k} = (k_1,k_2, \ldots, k_d), $ where $ \forall s = 1,2,\ldots \ k_s \ge 1,$
$$
\phi(\vec{k}, \vec{x}) = \prod_{s=1}^d \psi_{k_s}(x_s). \eqno(7)
$$
{\bf Remark 1.} Note that the functions $ \{ \phi_{\vec{k}}(\vec{x}) \} $ are eigen
functions for energy operator $ E_m. $ This circumstance  allow us a possibility to
generalize our method.\par
 Further, we introduce the sequence of real numbers $ \{ \lambda(k) \} $ as a
 square norms of  the functions $ \{ \psi_k(\cdot) \}: \ \lambda(k) =
 || d \psi_k(x)/dx||^2. $
It is easy to compute: $ \ \lambda(1)=0;$

$$
 \lambda(2) = \lambda(3)= 1; \ \lambda(4) = \lambda(5)= 4; \ \lambda(6)  = \lambda(7) = 9; \ \lambda(8)= \lambda(9)=16, \ldots.
$$
 Let us define
$$
\Lambda_d(k) = \Lambda_d( \vec{k} ) \stackrel{def}{=}
 \left[ \sum_{j=1}^d \lambda^2(k_j) \right]^{1/2}.
$$
 Note that

 $$
 \Lambda_d(k) \asymp \left( \sum_{j=1}^d k_j^2 \right)^{1/2}.
 $$

 The estimating function $ E_m f(\cdot) $ belongs to the space
$ L_2(-\pi,\pi)^d $ or equally $ W_m(f) < \infty $ if and only if

$$
W_m(f) = \sum_{\vec{k} }c^2(\vec{k}) \cdot \left[\Lambda_d(k) \right]^{2m}
< \infty. \eqno(8)
$$

{\bf Remark 2.} The expression (8) has a sense still in the case when the number $ m $
is non \ - \ integer.Further we will use only the expression (8) for the value $ W_m(f) $
tacking into account all the values $ m, \ m \in [0, \infty). $ \par

 We denote in the considered case $ W_m(f) < \infty $

 $$
 \rho_m(N) = \rho_m(f,N) = \sum_{\vec{k} \in R_3(N) }c^2(\vec{k}) \cdot
 \left[\Lambda_d(k) \right]^{2m}.
  $$

It is evident that $ \rho_m(N) \downarrow 0 $ as $ N \uparrow \infty. $\par

 The values $ \rho_m(N) = \rho_m(f,N) $ are known and well studied in the
{\it approximation theory, } see, e.g., ([9]). Namely, the variable $ \rho_0^{1/2}(N) $
is the error in the $ L_2[-\pi, \pi]^d $ norm of the best approximation
of a function $ f $ by means of  $ d \ - $ dimensional trigonometric polynomials
with the maximal degree less or equal than $ N: $

$$
\rho_0(N) = \inf_{T(N)} ||f \ - \ T(N)||^2,
$$
where  $ T(N) $ is the  $ d \ - $ dimensional trigonometric polynomials
with the maximal degree less or equal than $ N; $ and are closely connected with
a so \ - \ called Zygmund's $ L_2 $ modules of continuity:

$$
\omega_2(f, \delta) = \sup_{ |\vec{h}| \le \delta} ||f(\vec{x} + \vec{h}) \ - \ 2
f( \vec{x}) +  f(\vec{x} \ - \ \vec{h})||,  \ \delta \in [0,1],
$$
where the algebraic operations $ \vec{x} \pm \vec{h} $ are understood coordinatewise
 modulo $ 2 \pi.$ \par
 Indeed, the following implication holds:

 $$
 \rho_0^{1/2}(f,N) \asymp N^{- l - \nu}, \ l = 0,1,2 \ldots, \nu \in [0,1) \
 \Longleftrightarrow
 $$

$$
\omega_2(f^{(l)}, \delta) \asymp \delta^{\nu}.
$$

 We impose the following condition on the signal $ f(\cdot): \ $ for some $ \beta > 0, $

$$
\rho_m(f,N) \le C N^{-2 \beta}, \eqno(9)
$$
or more general condition:

$$
\overline{\lim}_{N \to \infty} \frac{\rho_m(2N) }{\rho(N) } < 1.
$$

 This conditions are satisfied in many practical cases.\par
 Further, we will consider only the {\it infinite \ - \ dimensional } case, i.e. when
 $ \forall N > 0 \ \Rightarrow \rho_m(N) > 0. $ Moreover, we suppose

 $$
 \exists h \in (0,1), \forall N \ge 3 \ \Rightarrow \rho_m(f,N) \ge C \ h^N. \eqno(10)
 $$

 As usually, we will denote $ A(N) \asymp B(N), N = 1,2,\ldots $ iff
$$
0 < C_1 = \inf_{N \ge 1} A(N)/B(N) \le \sup_{N \ge 1} A(N)/B(N) = C_2 < \infty.
$$

\vspace{3mm}

{\bf 2. Construction of an estimations.  }\\

\vspace{3mm}

 We estimate the unknown coefficients $ c(k)= c(\vec{k} ) $ as follows:

$$
c(\vec{k},n)  = n^{-1} \sum_{i=1}^n y(i) \ \phi( \vec{k}, \vec{x}_i). \eqno(11)
$$

The consistent estimation $ \sigma^2(n) $ of the variance $ \sigma^2 $ is offered in
[1]:

$$
\sigma^2(n) = n^{-1} \sum_{i=1}^n  \left(y(i) \ - \ f_n^{(0)}(x(i) \right)^2,  \eqno(12)
$$
where $  f_n^{(0)}(x) $ is any preliminary consistent estimation of the function
$ f(\cdot). $ \par

Let us offer a following {\it family } of a "projection"  estimations $ W_m(f,n,N) $
of the values $ W_m(f) $ of a view:

$$
{\bf W_m(f,n,N) = \sum_{\vec{k} \in R(N)}\left[\Lambda_d(k) \right]^{2m} \cdot
\left( c^2(n, \vec{k}) \ - \ n^{-1} \sigma^2(n)   \right). } \eqno(13)
$$

{\it We obtain after some calculations under conditions (9) and the following
condition (14):  }

$$
||f||S(\max(d,2m)) < \infty: \eqno(14)
$$

$$
{\bf Z(W_m(f,n,N),  W_m(f)) \asymp n^{-1} + \rho_m^2(N) + n^{-2} N^{4m + d}.} \eqno(15)
$$

\vspace{3mm}

{\bf 3. Non \ - \ adaptive estimation.}\\

\vspace{3mm}

 We suppose that the conditions (9) and (15) are satisfied. Tacking the value $ N_0 $
as follows:

$$
N_0 = \argmin_N \left( N^{-2\beta} + n^{-1} N^{2m + d/2} \right),
$$
i.e.

$$
N_0 \asymp \left[ n^{1/(2 \beta + 2m + d/2 ) } \right] \stackrel{def}{=} N_0(\beta),
$$
we obtain:

$$
 Z(W_m(f,n,N_0), W_m(f)) \le K(\beta,m,d) \
\max \left[ n^{-1}, n^{-4\beta/( 2 \beta + 2 m + d/2 ) } \right].
\eqno(16)
$$

 Note that in the case

 $$
 \beta \ge m + d/4 \eqno(17)
 $$

$$
{\bf Z(W_m(f,n,N_0), W_m)  \le K_1(\beta,m,d) \ n^{-1}. } \eqno(18)
$$

 The inequality (18) show us that the estimation $ W_m(f, n,N_0) $ is asymptotically
 optimal. But it is non \ - \ adaptive, since it dependent on the parameter $ \beta, $
 which is unknown as usually in many practical cases. \par
  In the next section we offer the so \ - \ called adaptive estimation, i.e. which
  dependent only on the observations $ \{ y(i) \} $ and has again the optimal speed
  of convergence, but under more strictly conditions.\par

  \vspace{3mm}

 {\bf 4. Adaptive estimation.}\\

  \vspace{3mm}

Let us introduce the following important functional:

$$
\tau(N) = \tau(N,n) = \sum_{\vec{k}\in R_2(N) } c^2(\vec{k},n ) \
 [\Lambda_d( \vec{k} ) ]^{2m}, \eqno(19)
$$
which dependent only on the source data $ \{ y(i) \} $, and let us choose the following
(random) value of harmonics:

$$
{\bf \hat{N} = \argmin_{ N \le N_{d,m}(n) } \tau(N,n). } \eqno(20)
$$

 It may be proved  that under condition (9)
 as in articles [3], [4] in the sense of convergence with
 probability one and in the $ L_2(\Omega) $ norm

 $$
 \lim_{n \to \infty} \hat{N} /n^{1/( 2 \beta + 2m +d)  } = K_3(\beta,m,d), \eqno(21)
 $$
where $ K_3 = K_3(\beta ,m,d) $ is some positive finite non \ - \ random constant.\par

  Correspondingly in the choice of $ N_0, $ we can offer the following still adaptive
  estimation of $ W_m(f): $

  $$
 {\bf \hat{W}_m(f,n) = W_m(f,n, \hat{N}).} \eqno(22)
  $$

 Substituting into the expression for the loss function $ Z(\cdot, \cdot) $ and using
  the equality (21), we conclude that under conditions (9), (14)

$$
{\bf Z(\hat{W}_m(f,n), W_m) \asymp \max
\left( n^{-1}, n^{-4\beta/(2\beta + 2 m + d) } \right). } \eqno(23)
$$

 Notice that in the case when

 $$
 \beta \ge m + d/2 \eqno(24)
 $$
 (c.f.  with the condition (17)), the adaptive estimation (22) $ \hat{W}_m(f,n) $ is
asymptotically  optimal:

$$
{\bf Z(\hat{W}_m(f,n), W_m) \le K_4(\beta, m,d) \ n^{-1}. } \eqno(25)
$$

\vspace{3mm}

{\bf 5. Concluding remark.}\\

\vspace{3mm}

We conclude as in the case density energy estimations [11], [12]
still under condition  \par

$$
W_{max(d,2m) }(f) < \infty:
$$

{\bf A.} In the case when $ \beta < m + d/4 $ the optimal  estimation of $ W_m(f) $
is impossible.\par

\vspace{2mm}

{\bf B.} In the case

$$
m + d/4 \le \beta < m + d/2
$$

there exists only the non \ - \ adaptive estimation with optimal rate of convergence
$ n^{-1/2}.$ \par

\vspace{2mm}

{\bf C.} If ultimately

$$
\beta \ge m + d/2,
$$
then there exists the optimal adaptive estimation for the energy value $ W_m.$ \par
 The optimality denotes, by our definition, that the speed of of convergence is
 (asymptotically) $ 1/\sqrt{n}. $ \par

 \vspace{2mm}

{\bf D.} Note in addition that all the offered in this report estimations
$ W_m(f,n,N) $ are asymptotically  normal distributed with parameters
correspondingly

 $$
 {\bf E} W_m(f,n,N) \ - \ W_m \sim \rho_m(N),
 $$

$$
{\bf Var} \left(  W_m(f,n,N)  \right) \asymp n^{-1} \ ||f||S(\max(d,2m))
 + \rho^2_m(N) + N^{4m + d}/n^2.
$$

 The last circumstance may be used by building of confidence interval for $ W_m(f)$
 and for verifications of different statistical hypothesis.\par

\vspace{2mm}

{\bf  F.} Evidently, we can use  by the computing of the offered here estimations
the famous Fast Fourier Transform (FFT) with the computation of complexity $ \asymp
C(d) \ n \ \log n. $ \par

\vspace{2mm}

{\bf E.} The accuracy proof of our assertions and building of the confidence region
for $ W_m(f) $ is at the same as in the articles [3] \ - \ [5]. It used the approximation theory, theory of martingales, for instance, the exponential bounds in the Law of
Iterated Logarithm (LIL), theory of Banach spaces of random variables etc.\par

\vspace{2mm}

\vspace{4mm}

\begin{center}

            {\bf  References}\\

 \end{center}

1. Golubev G., Nussbaum M.  {\it Adaptive spline Estimations in the nonparametric regression Model.}  Theory Probab. Appl., 1992, v. 37 $ N^o $ 4, 521 - 529.\\
2. Juditsky A., Nemirovsky A. {\it Nonparametric denoising Signals of Unknown
Local Structure, II: Nonparametric Regression Estimation.} Electronic Publication, arXiv:0903.0913v1 [math.ST] 5Mar2009.\\
3. Ostrovsky E., Sirota L. {\it Universal adaptive estimations and confidence  intervals in the non-parametrical statistics.} Electronic Publications, arXiv.mathPR/0406535 v1 25 Jun 2004.\\
4. Ostrovsky E., Zelikov Yu. {\it Adaptive Optimal Nonparametric Regression and Density Estimation based on Fourier \ - \ Legendre
Expansion. } Electronic Publication, arXiv:0706.0881v1 [math.ST]
6 Jun 2007. \\
5. Ostrovsky E., Sirota L. {\it Optimal adaptive nonparametric denoising of
Multidimensional \ - \ time signal.} Electronic Publication,
arXiv:0809.30211v1 [physics.data\ - \ an] 17 Sep 2008.
6. Donoho D.  {\it Wedgelets: nearly minimax estimation of edges.}  Annals of  of Statist., 1999, v. 27 b. 3 pp. 859 - 897.\\
7. Donoho D.  {\it Unconditional bases are optimal bases for data compression and for statistical estimation.} Applied Comput. Harmon. Anal., 1996, v. 3 pp. 100 - 115.\\
8.Keipers R., Niederreiter W. {\it The uniform Distribution of the
Sequences. } Kluvner Verlag, Dorderecht, 1983.\\
9. Dai F., Ditzian Z., Tikhonov S. {\it Sharp Jackson inequalities.} Journal
of Approximation Theory. 2007, doi:10.1016/j.jat.2007, 04.015.\\
10. Noullez A., Vergassola M. {\it A fast Legendre transform algorithm and
Applications to the adhesion model.} Journal of Scientific Computing, Springer,
1994, v. 9, $ N^o 3, $ p. 259 \ - \ 281.\\
11. Efromovich S. and Low M. {\it On Bickel and Ritov conjecture about adaptive
estimation of the integral of the square of density derivative.} The Annals of Statistics, 1996, Vol. 24, No 2, 682 \ - \ 686.\\
12.Bickel P.J., and Ritov Y. {\it Estimating integrated squared density derivatives : sharp best order of convergence estimates.} Sankhya, Ser. 1988, {\bf A 50}, 381 \ - \ 393.\\
13. Donoho D. and Nussbaum M. {\it Minimax quadratic estimation of a quadratic
functional}. J. Compexity , 1990, {\bf 6}, 290 \ - \ 323.\\
14. Efromovich S.{\it Nonparametric estimation of a density with unknown smoothness.}             Theory Probab. Appl., 1985,  {\bf 30}, 557 \ - \ 568.\\
15. Fan J. {\it On the estimation of quadratic functionals}. Annals Statist., 1991,
{\bf 19}, 1273 \ - \ 1294.\\

\end{document}